\documentclass[Physsubmission, Phys]{SciPost}

\hypersetup{
    colorlinks,
    linkcolor={red!50!black},
    citecolor={blue!50!black},
    urlcolor={blue!80!black}
}

\usepackage[bitstream-charter]{mathdesign}
\DeclareSymbolFont{usualmathcal}{OMS}{cmsy}{m}{n}
\DeclareSymbolFontAlphabet{\mathcal}{usualmathcal}

\usepackage{amsmath}
\usepackage{bm}
\usepackage{epsfig}
\usepackage{graphicx}
\usepackage{listings}
\usepackage{ragged2e}
\usepackage{subcaption}
\usepackage{fancyvrb}	

\usepackage{tikz}
\usetikzlibrary{arrows}
\usetikzlibrary{decorations.markings}
\usetikzlibrary{decorations.pathmorphing}
\usetikzlibrary{decorations.pathreplacing}
\usetikzlibrary{shapes}
\usetikzlibrary{trees}

\input{all.tikzdefs}

\tikzstyle{blank}=[fill=white, shape=circle, draw=white, inner sep=1.0pt]
\tikzstyle{dot}=[fill=black, shape=circle, draw=black, inner sep=1.0pt]

\tikzstyle{line}=[-, draw=black!80!white, line width=1pt, line cap=rect]
\tikzstyle{thick line}=[-, draw=black, line width=1.5pt, line cap=rect]
\tikzstyle{thin line}=[-, draw=black, line width=0.6pt, line cap=rect]
\tikzstyle{external edge}=[-, line width=1pt, line cap=rect, draw={rgb,255: red,102; green,102; blue,102}]
\tikzstyle{massless external edge}=[-, line width=0.8pt, densely dashed, line cap=rect, draw={rgb,255: red,102; green,102; blue,102}]
\tikzstyle{edge}=[-, draw={rgb,255: red,176; green,36; blue,39}, line width=1pt, preaction={{draw=white,line width=2pt}}, line cap=rect]
\tikzstyle{massless edge}=[-, draw={rgb,255: red,176; green,36; blue,39}, line width=0.8pt, densely dashed, line cap=rect]
\tikzstyle{dot1}=[-, postaction=decorate, decoration={markings,mark=at position .50 with {\node[style=dot]{};}}]
\tikzstyle{dot2}=[-, postaction=decorate, decoration={markings,mark=between positions 0.33 and 0.67 step 0.33 with {\node[style=dot]{};}}]
\tikzstyle{dot3}=[-, postaction=decorate, decoration={markings,mark=between positions 0.25 and 0.76 step 0.25 with {\node[style=dot]{};}}]
\tikzstyle{dot4}=[-, postaction=decorate, decoration={markings,mark=between positions 0.20 and 0.81 step 0.20 with {\node[style=dot]{};}}]

\usepackage{booktabs}
\usepackage{array}
\newcolumntype{M}[1]{>{\centering\arraybackslash}m{#1}}	

\usepackage[super]{nth}
\definecolor{light-gray}{gray}{0.97}

\usepackage[colorinlistoftodos, textsize=scriptsize]{todonotes}
\graphicspath{{plots/}}
\newcounter{bla}

\def\be{\begin{equation}}
\def\ee{\end{equation}}
\def\bea{\begin{align}}
\def\eea{\end{align}}
\def\nn{\nonumber}

\newcommand{\pysecdec}{py{\textsc{SecDec}}}

\newcommand{\form}{{\textsc{Form}}}

\newcommand{\eps}{\varepsilon}

\begin{document}

\begin{center}{\Large \textbf{
\pysecdec{} as a tool for expansion by regions and loop amplitude evaluation\\
}}\end{center}

\begin{center}
Emilio Villa\textsuperscript{1},
\end{center}

\begin{center}
{\bf 1} Institute for Theoretical Physics, Karlsruhe Institute of Technology (KIT),\\ 76131 Karlsruhe, Germany
\\
* emilio.villa@kit.edu
\end{center}

\begin{center}
\today
\end{center}


\definecolor{palegray}{gray}{0.95}
\begin{center}
\colorbox{palegray}{
  \begin{tabular}{rr}
  \begin{minipage}{0.1\textwidth}
    \includegraphics[width=35mm]{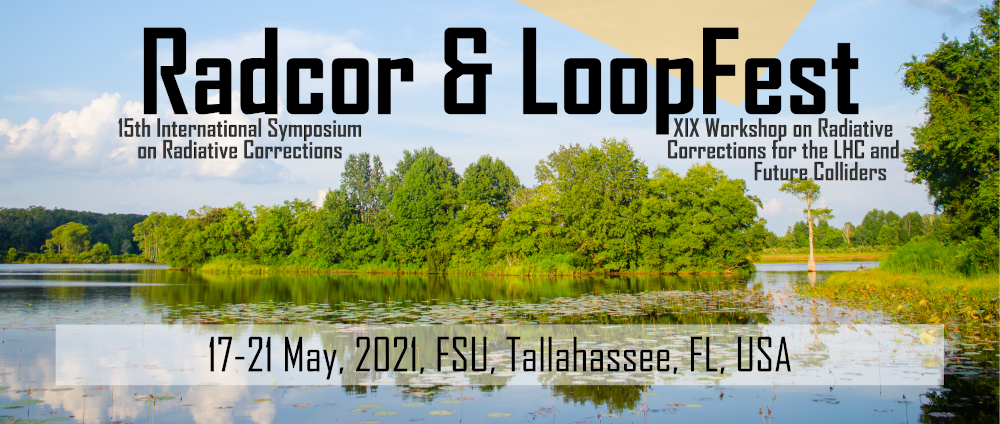}
  \end{minipage}
  &
  \begin{minipage}{0.85\textwidth}
    \begin{center}
    {\it 15th International Symposium on Radiative Corrections: \\Applications of Quantum Field Theory to Phenomenology,}\\
    {\it FSU, Tallahasse, FL, USA, 17-21 May 2021} \\
    \doi{10.21468/SciPostPhysProc.?}\\
    \end{center}
  \end{minipage}
\end{tabular}
}
\end{center}

\section*{Abstract}
{\bf
We present the new release of \pysecdec{}, a toolbox for the
evaluation of dimensionally regulated parameter integrals. The main
new features consist of an automated way to perform expansions, based
on the geometric approach to the method of expansion by regions, and a
new algorithm to efficiently evaluate linear combinations of integrals
as needed for the calculation of scattering amplitudes. The other new features are also summarised briefly.
}


\section{Introduction}
\label{sec:intro}

The importance of precision calculations for the current and future High Energy Physics program is undeniable,
however their complexity increases rapidly with the numer of loops and mass scales.
This often prohibits analytic calculations of the underlying scattering amplitudes, such that (semi-)numerical approaches need to be pursued.
Furthermore, in certain limits leading to a hierarchy of scales, approximate results can still be accurate enough for phenomenological purposes.
This has been exploited for example in Higgs phenomenology, exploiting large-$m_t$ or small-$m_b$ expansions in their respective ranges of validity, see e.g. Ref.~\cite{Heinrich:2020ybq} for more details.

A systematic approach to the expansion of Feynman integrals is given by the so-called {\em expansion by regions}, pioneered in Refs.~\cite{Smirnov:1991jn,Beneke:1997zp,Smirnov:1998vk}. The method, originally developed in the momentum representation of Feynman integrals, was later reformulated in Feynman parameter space~\cite{Smirnov:1999bza,Pak:2010pt,Ananthanarayan:2018tog,Ananthanarayan:2020ptw}, where it allows for a geometric interpretation.
An implementation of the method is available in the code \textsc{asy2.m}~\cite{Jantzen:2012mw}, which is also part of the program \textsc{Fiesta}~\cite{Smirnov:2008py,Smirnov:2009pb,Smirnov:2013eza,Smirnov:2015mct,Smirnov:2021rhf}.

In these proceedings, we present a new version~\cite{Heinrich:2021dbf} of \pysecdec~\cite{Borowka:2017idc,Borowka:2018goh}, which contains an implementation of expansion by regions based on its geometric formulation.
Moreover, we present a new mechanism to efficiently evaluate amplitudes as sums of integrals, in a way which takes into account the relative importance of the individual integrals. Furthermore we briefly introduce the other features of the new release, for more details we refer to~~\cite{Heinrich:2021dbf}.

\section{Geometric formulation of expansion by regions}
\label{sec:1}

\subsection{Method}
\label{subsec:11}

In this section, we discuss the expansion of parametric integrals over polynomials around small values of a so-called {\em smallness parameter} (e.g. $m^2/p^2$ in a large-momentum expansion).
Feynman integrals can be brought into this form using the Lee-Pomeransky representation~\cite{Lee:2013hzt}:
\begin{equation}
I = \int_0^{\infty}\,\frac{\mathrm{d}\mathbf{x}}{\mathbf{x}} \mathbf{x}^{\bm{\nu}} t^{\nu_{N+1}} \left[ \sum_{i=1}^m c_i \mathbf{x}^{\mathbf{p}_i} t^{p_{i,N+1}} \right]^{-\frac{D}{2}}\;,
\label{eq:int}
\end{equation}
where $t$ is the smallness parameter, $c_{i}$ the coefficients of the polynomial, $D$ the space-time dimension and $\mathbf{x}^{\mathbf{a}} =\prod_{j=1}^N x_j^{a_j}$.
The exponents can be organised into $(N+1)$-dimensional vectors $\mathbf{p}'_i \equiv \left(\mathbf{p}_i, p_{i,N+1} \right)$, $\bm{\nu}' \equiv \left(\bm{\nu}, \nu_{N+1} \right)$.

The goal of the method of expansion by regions is to identify all the possible scalings of the integration variables w.r.t. the smallness parameter in the different parts (called \textit{regions}) of the integration domain. Once the integration domain has been divided into regions, the integrand is expanded in each region according to the given scaling and the corresponding expanded integrands are integrated over the \textit{whole} integration domain. The sum of the resulting integrals adds up to the original integral since overlapping terms in general give rise to scaleless integrals that are zero in dimensional regularisation~\cite{Jantzen:2011nz}. Moreover, the dimensional regulator $\eps$, or additional analytical regulators, are used to make sure the integrals are well-defined even outside the region of convergence.
Let us now sketch the geometric approach and illustrate it with a simple example.

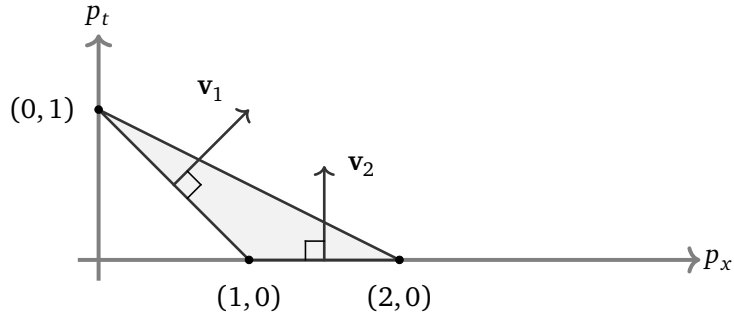
\begin{figure}[h]
    \centering
    {\begin{tikzpicture}
	\begin{pgfonlayer}{nodelayer}
		\node [style=none] (0) at (0, -0.25) {};
		\node [style=none] (1) at (8, 0) {};
		\node [style=none] (2) at (0, 3) {};
		\node [style=dot] (3) at (0, 2) {};
		\node [style=dot] (4) at (2, 0) {};
		\node [style=dot] (5) at (4, 0) {};
		\node [style=none] (6) at (3, 0) {};
		\node [style=none] (7) at (1, 1) {};
		\node [style=none] (8) at (2, 2) {};
		\node [style=none] (9) at (3, 1.25) {};
		\node [style=none] (10) at (-0.75, 2) {$(0,1)$};
		\node [style=none] (11) at (2, -0.5) {$(1,0)$};
		\node [style=none] (12) at (4, -0.5) {$(2,0)$};
		\node [style=none] (13) at (0, 3.25) {$p_t$};
		\node [style=none] (14) at (8.25, 0) {$p_x$};
		\node [style=none] (15) at (3.5, 1.25) {$\mathbf{v}_2$};
		\node [style=none] (16) at (1.5, 2.25) {$\mathbf{v}_1$};
		\node [style=none] (17) at (1.36, 1) {};
		\node [style=none] (18) at (1.18, 0.82) {};
		\node [style=none] (19) at (1.18, 1.18) {};
		\node [style=none] (20) at (2.75, 0) {};
		\node [style=none] (21) at (2.75, 0.25) {};
		\node [style=none] (22) at (3, 0.25) {};
		\node [style=none] (23) at (-0.25, 0) {};
	\end{pgfonlayer}
	\begin{pgfonlayer}{edgelayer}
		\draw [style=thick line, ->, gray] (0.center) to (2.center);
		\draw [style=thick line, ->, gray] (23.center) to (1.center);
		\draw [style=line, line join=round, fill={black!5!white}] (3.center)
			 to (4.center)
			 to (5.center)
			 to cycle;
		\draw [style=line, ->] (6.center) to (9.center);
		\draw [style=line, ->] (7.center) to (8.center);
		\draw [style=thin line] (19.center) to (17.center);
		\draw [style=thin line] (17.center) to (18.center);
		\draw [style=thin line] (20.center) to (21.center);
		\draw [style=thin line] (21.center) to (22.center);
	\end{pgfonlayer}
\end{tikzpicture}}
    \caption{Newton polytope for $P(x,t) = t+x+x^2$, together with the directions of the region vectors (drawn as normal vectors to the facets in positive $t$-direction).}
    \label{fig:polytope_simple}
\end{figure}

\noindent In order to find all the relevant regions, we can use the notion of the {\em Newton polytope} of a polynomial.
The Newton polytope can be determined as the convex hull of the exponent vectors or, alternatively, as the intersection of half spaces
\begin{equation}
\Delta' = \bigcap_{f\in F} \left\{ \mathbf{m}\in\mathbb{R}^{N+1} \mid \langle \mathbf{m},\mathbf{n}_f\rangle + a_f \geq 0 \right\}\;,
\end{equation}
where $F$ is the set of polytope facets with inward-pointing normal vectors $\mathbf{n}_f$, $\langle \mathbf{m},\mathbf{n}_f\rangle$ is the scalar product of $\mathbf{m}$ and $\mathbf{n}_f$, and $a_f\in \mathbb{Z}$.
An example of a Newton polytope for the  simple polynomial $ P(x,t) = t+x+x^2 $ is shown in Fig.~\ref{fig:polytope_simple}.

The subset of facets with normal vectors pointing in the positive $t$-exponent direction is $F^+ = \left\{ f \in F \mid (\mathbf{n}_f)_{N+1} > 0\right\}$. It can be shown that the facets belonging to $F^+$ correspond to all the regions we need to consider~\cite{Pak:2010pt}. These can be used as input for the change of variables:
\be
t \rightarrow \prod_{f \in F^+} z_f^{(\mathbf{n}_f)_{N+1}} t\;,~~~~~~~
x_i \rightarrow  \prod_{f \in F^+} z_f^{(\mathbf{n}_f)_i} x_i\;,
\label{eq:trafo}
\ee
where for each facet $f$, $\mathbf{n}_f$ is the vector describing the rescaling applied to $(\mathbf{x},t)$ and is called the {\em region vector}.

The transformations \eqref{eq:trafo} lead to the following form of the integral~\eqref{eq:int}:
\be
I = \left(\prod_{f\in F^+} z_f^{\langle \mathbf{n}_f,\bm{\nu}'\rangle + \frac{D}{2} a_f}\right) \int_0^{\infty}\,\frac{\mathrm{d}\mathbf{x}}{\mathbf{x}} \mathbf{x}^{\bm{\nu}} t^{\nu_{N+1}}  \left[ \sum_i c_i \mathbf{x}^{\mathbf{p}_i} t^{p_{i,N+1}} \prod_{f\in F^+} z_f^{\langle \mathbf{n}_f,\mathbf{p}'_i\rangle + a_f}  \right]^{-\frac{D}{2}}.
\label{eq:intrescaled}
\ee
The original integral can then be approximated by
expanding in $z_f$ in each region while setting the other $z_{f'}$ to one,
and then setting  $z_f$ to one after the expansion in region $f$.
The final result reads:
\be
I = \sum_{f\in F^+} I_f\;,
\label{eq:ebrgen}
\ee
where $I_f$ is the expansion of \eqref{eq:intrescaled} in $z_f$ with all $z$ set to one after the expansions.
The $I_{f}$ are integrated over the whole integration domain.
The procedure outlined above is equivalent to expanding directly in the parameter $t$ after appropriate rescaling of the Feynman parameters, a proof can be found in Ref.~\cite{Heinrich:2021dbf}.

\subsection{Usage}
\label{subsec:12}
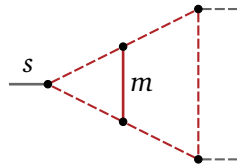
\begin{figure}[h]
\begin{center}
{\begin{tikzpicture}
	\begin{pgfonlayer}{nodelayer}
		\node [style=dot] (0) at (-1.75, 0) {};
		\node [style=dot] (1) at (-0.75, 0.5) {};
		\node [style=dot] (2) at (0.25, 1) {};
		\node [style=dot] (4) at (-0.75, -0.5) {};
		\node [style=dot] (5) at (0.25, -1) {};
		\node [style=none] (6) at (0.75, -1) {};
		\node [style=none] (7) at (-2.25, 0) {};
		\node [style=none] (8) at (0.75, 1) {};
		\node [style=none] (9) at (-0.5, 0) {$m$};
		\node [style=none] (10) at (-2, 0.25) {$s$};
	\end{pgfonlayer}
	\begin{pgfonlayer}{edgelayer}
		\draw [style=external edge] (7.center) to (0);
		\draw [style=massless external edge] (5) to (6.center);
		\draw [style=massless external edge] (2) to (8.center);
		\draw [style=massless edge] (0) to (1);
		\draw [style=massless edge] (1) to (2);
		\draw [style=massless edge] (2) to (5);
		\draw [style=massless edge] (5) to (4);
		\draw [style=massless edge] (4) to (0);
		\draw [style=edge] (1) to (4);
	\end{pgfonlayer}
\end{tikzpicture}} 
    \caption{Two-loop triangle with one internal massive line}
\label{fig:triangle2L}
\end{center}
\end{figure}

In this section we present a comparison between expansion by regions and standard \pysecdec{}. 
To this end, we evaluate as a sample integral the two-loop triangle given in Fig.~\ref{fig:triangle2L} at order $\eps^0$ in the $\eps$-expansion. We consider the limit $s \ll m^{2}$ and we expand to leading order in the smallness parameter.

\begin{figure}
\begin{center}
\includegraphics[width=0.99\textwidth]{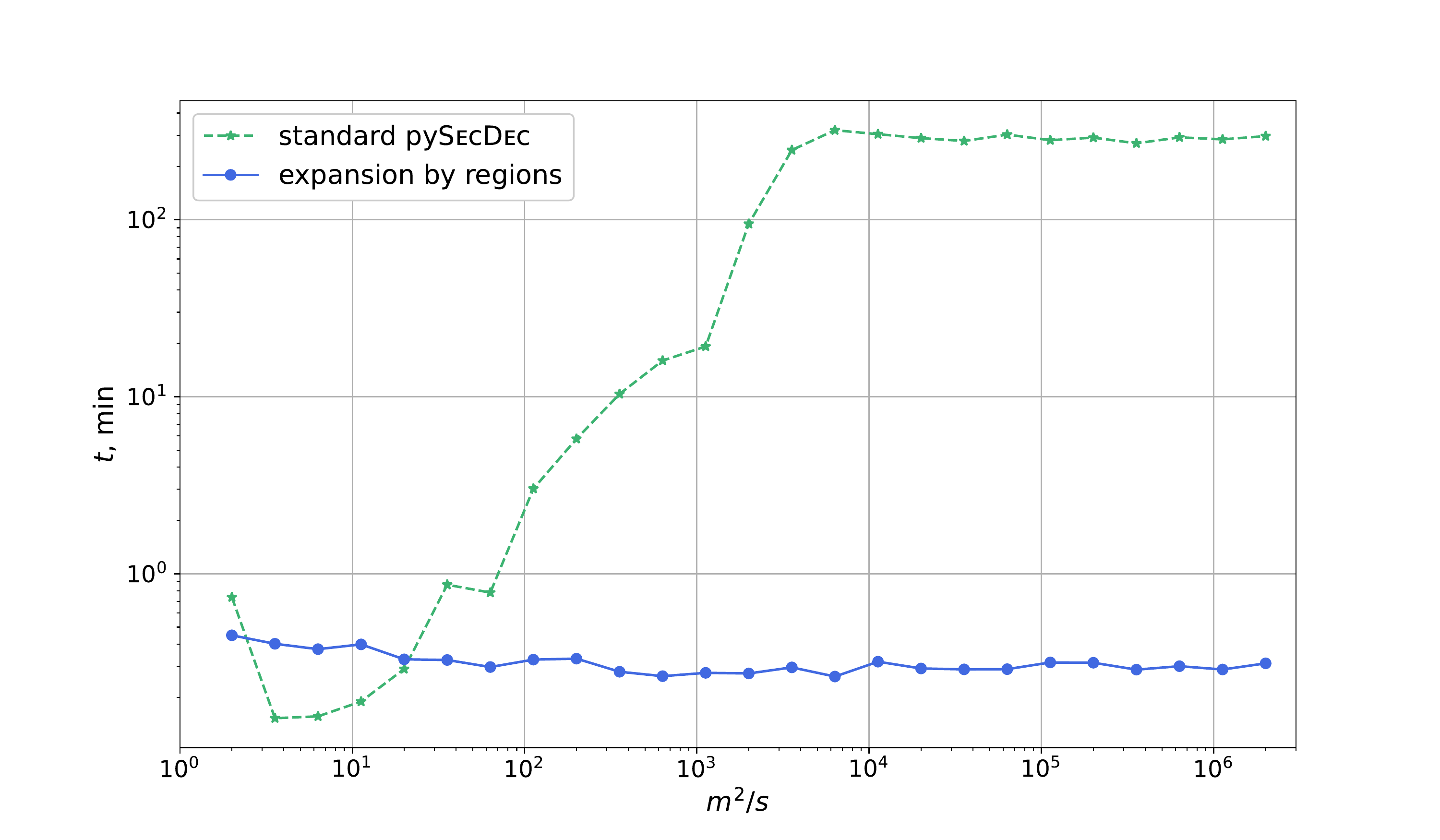} 
    \caption{Scan over different orders of magnitude of $r = m^2/s$ for the two-loop triangle given in Fig~\ref{fig:triangle2L}. Integration times (in minutes) are plotted against $r$. The relative accuracy goal is $10^{-3}$; the wall clock limit has been set to 5~hours.}
\label{fig:Scan}
\end{center}
\end{figure}

\noindent Fig.~\ref{fig:Scan} shows the integration times for both, expansion by regions (solid line) and \pysecdec{} (dashed line) plotted against the ratio $r=m^{2}/s$.
One can see how the integration times for \pysecdec{} blow up as the ratio of scales increases, while expansion by regions is numerically stable over the different orders of magnitude.

\begin{figure}
\begin{center}
\includegraphics[width=0.99\textwidth]{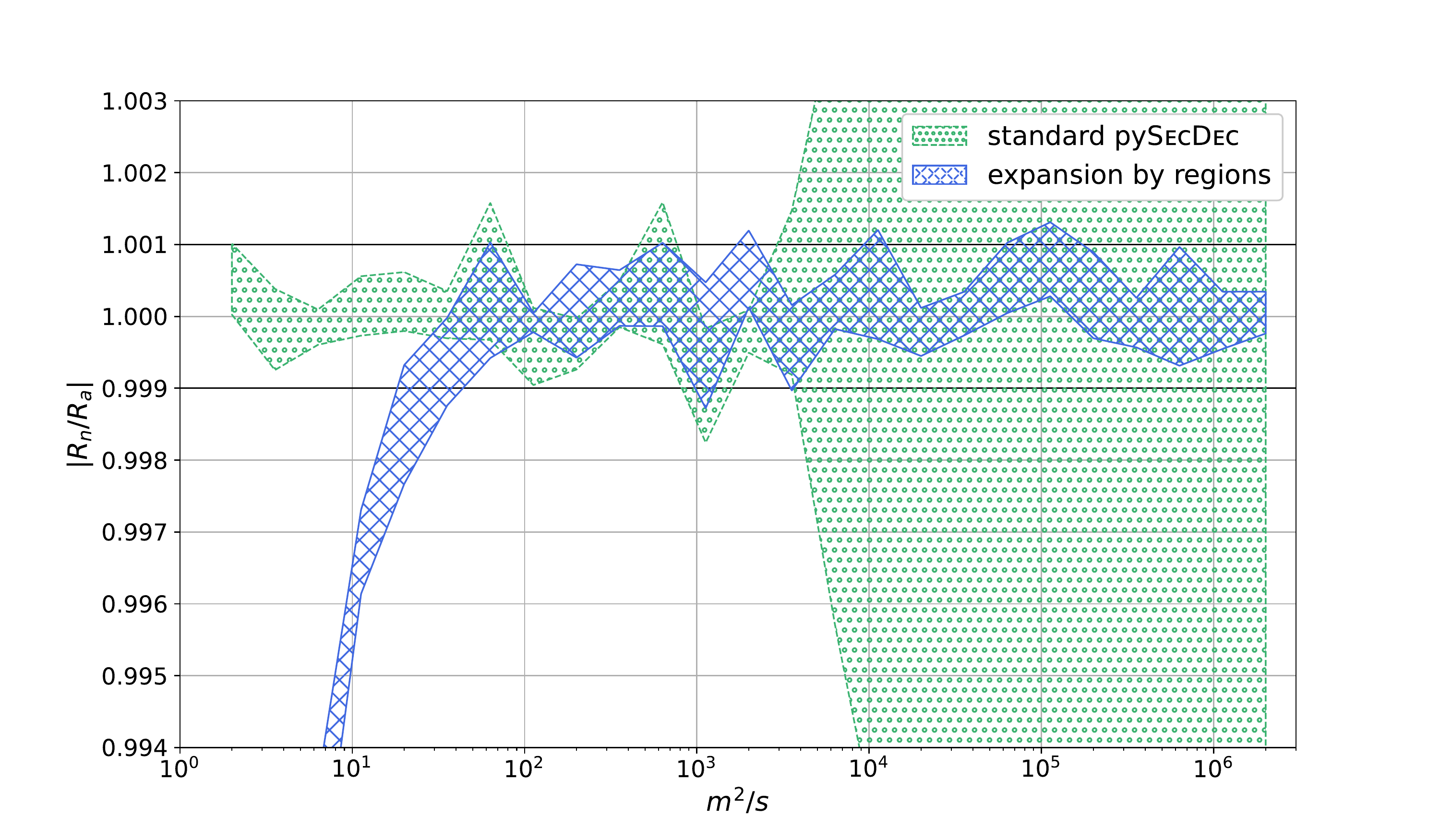} 
\caption{Scan over different orders of magnitude of $r = m^2/s$ for the two-loop triangle given in Fig~\ref{fig:triangle2L}. The modulus of the ratio between the numerical result and the analytic result, $|R_{n}/R_{a}|$, is plotted against $r$. The bands indicate the numerical uncertainties on the result.}
\label{fig:ScanAnalytic}
\end{center}
\end{figure}

Fig.~\ref{fig:ScanAnalytic} shows the behaviour of $|R_{n}/R_{a}|$, the modulus of the ratio of the finite parts obtained numerically to the analytic result (from Ref.~\cite{Fleischer:1998nb}), with increasing ratio $r = m^2/s$ of the kinematic scales, for both 
the numerical result obtained with expansion by regions and \pysecdec{}.
The shaded areas are given by adding and subtracting the numerical uncertainty to the result. The accuracy goal is fixed to $10^{-3}$.
Notice that the integration stops whenever the accuracy goal or the maximum number of evaluations are reached, meaning that the actual final uncertainty can be smaller than the required accuracy. 
Fig.~\ref{fig:ScanAnalytic} shows that the uncertainty due to the expansion is no longer dominant compared to the integration error if $m^2/s \gtrsim 40$, and that the results from \pysecdec{} and expansion by regions are compatible with each other in this case.

Combining the information of the two plots, it is clear that whenever the approximation due to the expansion is negligible compared to the numerical uncertainty of the result, using the expansion by regions option instead of standard \pysecdec{} is the better choice to evaluate the integral in the given kinematic limit. 

As a final remark, it is worth to point out that for some integrals, the above considerations might hold only much deeper in the kinematic limit.

\section{Amplitude evaluation and other new features}
\label{sec:2}

\subsection{Amplitude evaluation}
\label{subsec:21}
With the new release of \pysecdec{}, a new function called \texttt{sum\_package} has been introduced. The function implements the algorithm described in \cite{Kerner:2016msj} and generates a library that allows for the fast and efficient evaluation of amplitudes, i.e. linear combinations of integrals with coefficients depending on kinematic invariants and the space-time dimension.
The efficient evaluation is achieved by an algorithm that takes into account how much the individual integrals contribute to the total, such that the accuracy goal for less dominant integrals can be lower than the accuracy demanded for the total amplitude.
In this section we present a simple usage example, for more details we refer to Ref.~\cite{Heinrich:2021dbf}.

The example discussed here is contained in \texttt{examples/yyyy1L} of the public code and calculates the one-loop 4-photon amplitude 
${\cal M}^{++--}$. The amplitude can be expressed in terms of one-loop 4-point and 2-point integrals as
\begin{align}
{\cal M}^{++--} =-8 
 \left\{ 3(4-D) I_{4}^{D+4}(t,u) + \frac{t^{2}+u^{2}}{s} I_{4}^{D+2}(t,u) +  \frac{t-u}{s}\left(
    I_{2}^{D}(u)-I_{2}^{D}(t) \right) \right\}\,.
\label{eq:masters}
\end{align}
where $D=4-2\eps$  and $s,t,u$ are the usual Mandelstam invariants.
Note that $I_{4}^{D+4}$ is UV divergent, $I_{4}^{D+4}=1/(6\eps)+$finite,  and therefore provides the
rational part.

The example folder contains four files: 
\begin{itemize}
\item \texttt{coefficients.py}: here the coefficient functions of the integrals are defined as a list.
 Each coefficient function has the arguments \texttt{numerators,denominators,parameters}, where the field \texttt{parameters} contains the names of the kinematic invariants. The polynomials in the numerator and denominator can also depend on the regulator $\eps$.
\item \texttt{integrals.py}: here the integrals are given as a list containing the ``master'' integrals $I_{2}^{D}(u),~I_{2}^{D}(t),~
I_{4}^{D+2}(t,u),~I_{4}^{D+4}(t,u)$. Note that the ordering of the integrals and the corresponding coefficients should be the same.
\item \texttt{generate\_yyyy1L.py}: imports the integral definitions and the coefficients and runs \texttt{sum\_package}. 
\item \texttt{integrate\_yyyy1L.py}: performs the numerical integration, here the user can choose the integrator settings.
\end{itemize}

\noindent With the values of the kinematics invariants set to $t=-1.3, u=-0.8, s=-t-u$, the result reads
\begin{align}
  {\cal M}^{++--} &= (+0. \pm 2.1\cdot 10^{-16} )\,\eps^{-1} +  \nn\\
                  &+ (-28.431595834\pm 5.4\cdot 10^{-10}+
                          (-1.3\cdot 10^{-10} \pm 6.4\cdot 10^{-10}) \, i) \, + \nn\\
&+\mathcal{O}(\eps)\;,
\end{align}
in agreement with the analytic result. For more complicated amplitudes, the coefficients and the integral list will be much more complicated, however the structure is the same. Therefore \pysecdec{} can be used like an integral library for master integrals whose analytic expressions are unknown.

\subsection{Other new features}
\label{subsec:22}

In this section, we briefly introduce the other new features of \pysecdec{}. The following changes have been made compared to \pysecdec{} version 1.4.5:
\begin{itemize}

\item The contour deformation parameters $\lambda_i$ are now reduced automatically if the original values lead to an invalid contour.
  This removes the ``\textit{sign check error}'' that was one of the most frequent issues in previous versions of the code.

\item The \texttt{WorkSpace} parameter of \form{}~\cite{Kuipers:2013pba,Ruijl:2017dtg} is now automatically increased if \form{} fails due to insufficient \texttt{WorkSpace}. Users are no longer required to adjust the  \texttt{form\_work\_space} parameter.

\item \pysecdec{} can now be easily installed from the Python Package Index\footnote{\url{https://pypi.org/project/pySecDec/}} using
\begin{verbatim}
python3 -m pip install --user pySecDec 
\end{verbatim}

\item The functions \texttt{series\_to\_ginac}, \texttt{series\_to\_sympy}, \texttt{series\_to\_maple} and \\
\texttt{series\_to\_mathematica} have been added. They convert the output of \pysecdec{} to a syntax more suitable for use in combination with various computer algebra systems. 

\item The support for Python version~2.7 was dropped in favour of version~3.6 or newer. Python is now always invoked as \texttt{python3}.

\end{itemize}
The following new functions have been introduced:
\begin{itemize}
\item \texttt{make\_regions} provides a package generator to perform expansion by regions of general parameter integrals.
\item \texttt{loop\_regions} is a wrapper around \texttt{make\_regions} which simplifies the application of expansion by regions to loop integrals.
\item \texttt{sum\_package} has been discussed in the previous section.
\end{itemize}
These functions are documented in detail in the online documentation\footnote{\url{https://secdec.readthedocs.io}} distributed with the code.

\section{Conclusions}
In these proceedings, we have presented a new implementation of the method of regions in the program \pysecdec{}. We have shown that for integrals with a sufficiently large hierarchy among the kinematic invariants, the timings and accuracy of expansion by regions remain approximately constant as the ratio of scales increases, while the standard numerical evaluation by \pysecdec{} faces convergence issues.
This gives the user a powerful tool to get faster and more accurate results in certain kinematic limits.

In addition, we have introduced a mechanism in \pysecdec{}  that allows for the fast and efficient evaluation of linear combinations of integrals
with coefficients that depend on kinematics and regulators, typically occurring in multi-loop scattering amplitudes.
Each term in the sum is evaluated with a number of sampling points determined such that the global accuracy goal for the sum is reached most efficiently.
Together with the other features of the new release, this makes \pysecdec{} a tool suited for the evaluation of multi-loop amplitudes in a largely automated way.

\newpage

\section*{Acknowledgements}

I would like to thank the \pysecdec{} members for the fruitful collaboration.

\paragraph{Funding information}
This research was supported in part by the COST Action CA16201 (``Particleface'') of the European Union
and by  the  Deutsche  Forschungsgemeinschaft (DFG, German Research Foundation) under grant 396021762 - TRR 257.

\bibliography{refs.bib}

\nolinenumbers

\end{document}